\documentclass[reprint,NumberedRefs]{JASAnew}

\usepackage{amsthm}
\usepackage{booktabs}
\usepackage{mathtools}
\usepackage[ruled,vlined,linesnumbered,lined,boxed,noalgohanging]{algorithm2e}

\newcommand{\bom}[1]{\boldsymbol{#1}}
\newcommand{\bo}[1]{\mathbf{#1}}

\newcommand{\al}{\alpha} 
\newcommand{\be}{\beta}

\newcommand{\gam}{\gamma}  
\newcommand{\lam}{\lambda}

\newcommand{\delTh}{\Delta \theta}  

\newcommand{\beb}{\bom \beta}
\newcommand{\delb}{\bom \delta}
 
\newcommand{\vepsb}{\bom \varepsilon}

\newcommand{\thb}{\bom \theta} 

\newcommand{\Del}{\bom \Delta}

\newcommand{\bebh}{\hat{\beb}}

\newcommand{\mdim}{m}             
\newcommand{\ndim}{n}             
\newcommand{\pdim}{p}             
\newcommand{\kdim}{K}             

\renewcommand{\a}{\bo a}

\newcommand{\s}{\bo s}
 
\newcommand{\w}{\bo w}
\newcommand{\x}{\bo x}    
\newcommand{\y}{\bo y}

\newcommand{\A}{\bo A}

\newcommand{\X}{\bo X}   
\newcommand{\Xw}{\tilde \X}   

\newcommand{\hop}{\mathsf{H}} 
\newcommand{\im}{\jmath} 	   

\newcommand{\C}{\mathbb{C}}    
\newcommand{\R}{\mathbb{R}}    

\newcommand{\paino}{\sl}
\newcommand{\res}{\pmb{r}}  

\newcommand{\setA}{\mathcal A}
 
\newcommand{\ssgn}{\mathrm{sign}} 
\newcommand{\supp}{\mathrm{supp}}
 
\newcommand{\Xa}[1]{\bom \X_{\mathcal A_{#1}}}

\usepackage{array}
\newcolumntype{m}[1]{>{\centering\arraybackslash}p{#1}}

\newcommand{\beTKi}{\bebh^{\mathrm{init}}}

\newcommand{\beq}{\begin{equation}}
\newcommand{\eeq}{\end{equation}}
\newcommand{\bmat}{\begin{pmatrix}}
	\newcommand{\emat}{\end{pmatrix}}
\newcommand{\beqa}{\begin{eqnarray}}
\newcommand{\eeqa}{\end{eqnarray}}

\begin{document}

\title[Sequential adaptive elastic net]{Sequential adaptive elastic net approach for single-snapshot source localization}

\author{Muhammad Naveed Tabassum}
\author{Esa Ollila}

\affiliation{Aalto University, Dept. of Signal Processing and Acoustics, P.O. Box 15400, FI-00076 Aalto, Finland.}

\email{E-mail: (muhammad.tabassum, esa.ollila)@aalto.fi}

\begin{abstract}
This paper proposes efficient algorithms for accurate recovery of direction-of-arrivals (DoAs) of sources from single-snapshot measurements using compressed beamforming (CBF). In CBF, the conventional sensor array signal model is cast as an underdetermined complex-valued linear regression model and sparse signal recovery methods are used for solving the DoA finding problem. We develop a complex-valued pathwise weighted elastic net (c-PW-WEN) algorithm that finds solutions at knots of penalty parameter values over a path (or grid) of EN tuning parameter values. c-PW-WEN also  computes Lasso or weighted Lasso in its path. We then propose a sequential adaptive EN (SAEN) method that is based on c-PW-WEN algorithm with adaptive weights that depend on previous solution. Extensive simulation studies illustrate that SAEN improves the probability of exact recovery of true support compared to conventional sparse signal recovery  approaches such as Lasso, elastic net or orthogonal matching pursuit in  several challenging multiple target scenarios. The effectiveness of SAEN is more pronounced in the presence of high mutual coherence.
\end{abstract}

\maketitle

\section{Introduction}

Acoustic signal processing problems generally employ a system of linear equations as the data model. For overdetermined linear systems, the least square estimation (LSE) is often applied, but in underdetermined or ill-conditioned problems, the LSE is no longer unique but the optimization problem has infinite number of solutions. In these cases, additional constraints such as those promoting sparse solutions, are commonly used such as the Lasso (Least Absolute Shrinkage and Selection Operator) \cite{lasso:1996}  or the elastic net (EN) penalty \cite{zou2005reg} which is an extension of Lasso based on a convex combination of  $\ell_1$ and $\ell_2$ penalties of the Lasso and ridge regression. 

It is now a common practice in acoustic applications \cite{Stoica2008damas, xenaki2016block} to employ grid based sparse signal recovery methods for finding source parameters, e.g.,  direction-of-arrival (DoA) and power. 
This approach, referred to as  {\paino compressive beamforming (CBF)}, was originally proposed in  \cite{malioutov2005sparse} and has then  emerged as one of the most useful approaches in problems where only few measurements are available.
Since the pioneering work of \cite{malioutov2005sparse},  the usefulness of CBF approach has been shown in a series of papers \cite{Edelmann2011csbf, Gerstoft2014Compressive, Gerstoft2015Multiple, Seong2016spherical, Gerstoft2017Coherent, fortunati2014single,ollila:2015,ollila:2015b,tabassum2016single}.
In this paper, we address the problem of estimating the unknown source parameters when only a  single snapshot is available.

Existing approaches in grid-based single-snapshot CBF problem often use the Lasso for sparse recovery. Lasso,  however,  often performs poorly in the cases when sources are closely spaced in angular domain or when there exists large variations in the source powers. The same holds true when the grid used for constructing the array steering matrix is dense, which is the case, when one aims for high-resolution DoA finding. 
The problem is due to the fact that Lasso has poor performance when the predictors are highly correlated (cf.  \cite{zou2005reg, hastie2015stat}).  
Another problem is that Lasso lacks group selection ability. This means that when two sources are closely spaced in the angular domain, then Lasso tends to choose only one of them in the estimation grid but ignores the other.  EN  is often performing better in such cases as it enforces sparse solution but has a tendency to pick and reject the correlated variables as a group unlike the Lasso.   Furthermore, EN also enjoys the computational advantages of the Lasso \cite{zou2005reg} and adaptation using smartly chosen data-dependent weights can further enhance performance.

The main aim in this paper is to improve over the conventional  sparse signal recovery methods especially in the presence of high mutual coherence  of basis  vector  or  when non-zero coefficients  have largely varying amplitudes.   This former case occurs in CBF problem when a dense grid is used for constructing the steering matrix or when sources arrive to a sensor array from  either neighbouring or oblique angles. To this end, we propose a {\sl sequential adaptive approach} using the weighted elastic net (WEN) framework.
To achieve this in a computationally efficient way, we propose a homotopy method \cite{osborne2000new} that  is a complex-valued extension of the least angles regression and shrinkage (LARS) \cite{efron2004LAR} algorithm for weighted Lasso problem, which we refer to as c-LARS-WLasso.  
The developed c-LARS-WLasso method is numerically cost effective and avoids an exhaustive grid-search over candidate values of the penalty parameter. 

In this paper, we assume that the number of non-zero coefficients (i.e., number of sources $\kdim$ arriving at a sensor array in CBF problem) is known and propose a {\sl complex-valued pathwise (c-PW)-WEN} algorithm that utilizes c-LARS-WLasso along with PW-LARS-EN algorithm proposed in  \cite{tabassum2017pathwise} to compute the WEN path. c-PW-WEN computes the $\kdim$-sparse WEN solutions over a grid of EN tuning parameter values and then selects the best final WEN solution. 
We also propose a novel {\sl sequential adaptive elastic net (SAEN)} approach that applies adaptive  c-PW-WEN  sequentially by decreasing the sparsity level (order) from $3\kdim$ to $\kdim$ in three stages. 
SAEN utilizes smartly chosen adaptive (i.e., data dependent) weights that are  based on solutions obtained in the previous stage. 
  
Application of the developed algorithms is illustrated in the single-snapshot CBF DoA estimation problem.
In CBF, accurate recovery depends heavily on the user specified angular estimation grid  (the look directions of interests) which  determines the array steering matrix consisting of array response vectors to look directions (estimation grid points) of interests. A dense angular grid implies high mutual coherence, which indicates  a poor recovery region for most sparse recovery methods \cite{foucart2013mathematical}. 
Effectiveness of SAEN compared to  state-of-the art sparse signal recovery algorithms are illustrated  via extensive simulation studies. 

The paper is structured as follows. Section~\ref{sec:wen} discusses the WEN optimization problem and  its benefits over the Lasso and adaptive Lasso. 
We introduce  c-LARS-WLasso method in Section~\ref{sec:clarsen}.
In Section~\ref{sec:pwen}, we develop the c-PW-WEN algorithm that finds the $\kdim$-sparse WEN solutions over a grid of EN tuning parameter values.
In Section~\ref{sec:saen}, the SAEN approach is proposed.  
Section~\ref{sec:cbf} layouts the DoA's estimation problem from single-snapshot measurements using CBF. The simulation studies using large variety of  set-ups are provided in Section~\ref{sec:res}. Finally,    Section~\ref{sec:concl} concludes the paper.

{\it Notations}: Lowercase boldface letters are used for vectors  and uppercase for matrices. 
The  $\ell_2$-norm and    the $\ell_1$-norm are defined as $\| \a \|_2 = \sqrt{\a^\hop \a}$  and $\| \a\|_1 = \sum_{i=1}^\ndim |a_i |$,  respectively, where $| a | = \sqrt{a^* a} = \sqrt{a_R^2 + a_I^2} $ denotes the modulus of a complex number $a=a_R + \im a_I$.  
The support $\setA$ of a vector $\a \in \C^\pdim$ is the index set of its nonzero elements, i.e., $\setA = \supp(\a)= \{ j \in \{ 1, \ldots, \pdim\} : a_j \neq 0 \} $.  
The $\ell_0$-(pseudo)norm of  $\a$ is defined as $\| \a\|_0 = | \supp(\a)|$, which is equal to the total number of nonzero elements in it. 
For a vector $\beb \in \mathbb{C}^\pdim$ (resp. matrix $\X \in \mathbb{C}^{\ndim \times \pdim}$) and an index set $\setA_\kdim \subseteq \{1,2,\ldots, \pdim\}$ of cardinality $|\setA_\kdim|=\kdim$, we denote by $\beb_{\setA_\kdim}$ (resp.  $\X_{\setA_\kdim}$)  the $\kdim \times 1$ vector (resp. $\ndim \times \kdim$ matrix)   
restricted to components of $\beb$ (resp. columns of $\X$) indexed by the set $\setA_\kdim $. Let  $ \bo a \otimes \bo b$ denotes the Hadamard (i.e., element-wise) product  of $\bo a \in \C^\pdim$ and $\bo b \in \C^\pdim$ and by $\bo a \oslash \bo b$ we denote the element-wise division of vectors. 
We denote by $\langle \a, \bo b \rangle = \a^\hop \bo b$ the usual Hermitian inner product of  $\C^{\pdim}$.  Finally, 
$\mathrm{diag}(\a)$ denotes a $\pdim \times \pdim$ matrix with elements of $\bo a$ as its diagonal elements.

\section{Weighted Elastic Net Framework} \label{sec:wen}

We consider the linear model, 
where the $n$ complex-valued measurements are modeled as
\beq\label{eq:linear} 
\y = \X \beb + \vepsb,
\eeq
where $\X\in\C^{n\times p}$ is a known complex-valued design or measurement matrix, $\beb \in\C^{p}$ is the  unknown vector of complex-valued regression coefficients 
and $\vepsb \in\mathbb{C}^{n}$ is the complex noise vector.
For ease of exposition, we consider the centered linear model (i.e., we assume that the intercept is equal to zero).
In this paper, we deal with  underdetermined  or ill-posed linear model, where $\pdim > \ndim$, and the primary interest is to find a sparse estimate of the unknown parameters $\beb$ given $\y  \in\C^{n}$  and $\X\in\C^{n\times p}$. In this paper, we assume that the sparsity level, $\kdim= \|\beb \|_0$, i.e., the number of non-zero elements of $\beb$ is known. In DoA finding problem using compressed beamforming, this is equivalent to assuming that the number of sources arriving at a sensor array is known.

The WEN estimator finds the solution $\bebh  \in \mathbb{C}^\pdim$ to the following constrained optimization problem 
\beq\label{eq:cwen}  
\underset{\beb\in \mathbb{C}^\pdim}{\mathrm{minimize}}  \ \frac  1 2  \|\y - \X \beb \|_2^2 \quad \mbox{subject to}  \quad P_\al \bigl(\beb ; \w \bigr) \leq t, 
\eeq
where $t\geq0$  is the threshold parameter chosen by the user
and 
\[
P_\al\bigl(\beb;\w \bigr) = \sum_{j=1}^{p} w_j \Big(\al|\beta_j| + \cfrac{(1-\al)}{2}|\beta_j|^2\Big)
\]
is the WEN constraint (or penalty) function,  vector $\w=(w_1,\ldots,w_\pdim)^\top$, $w_i \geq 0$ for $i=1,\ldots, p$, collects the non-negative weights,  
and $\al\in[0, 1]$ is an {\paino EN tuning parameter}. Both the weights $\w$ and $\al$ are chosen by the user. 
When the weights are data-dependent, we refer the solution as {\sl adaptive EN (AEN)}. Note that AEN is an extension of the adaptive Lasso \cite{zou2006ALasso}.

The constrained optimization problem in \eqref{eq:cwen} can also be written 
in an equivalent penalized form  
\beq\label{eq:pwen} 
\hat \beb(\lam,\al) = \underset{\beb \in \C^\pdim}{\arg \min}  
\ \frac  1 2  \|\y - \X \beb \|_2^2   + \lam P_\al\bigl(\beb;\w\bigr),
\eeq 
where $\lam \geq 0$ is the {\paino penalty (or regularization) parameter}. The problems in  \eqref{eq:cwen} and  \eqref{eq:pwen} are equivalent  due to Lagrangian duality and either can be solved. Herein, we use \eqref{eq:pwen}. 

\begin{figure}[h]
\figline{
\fig{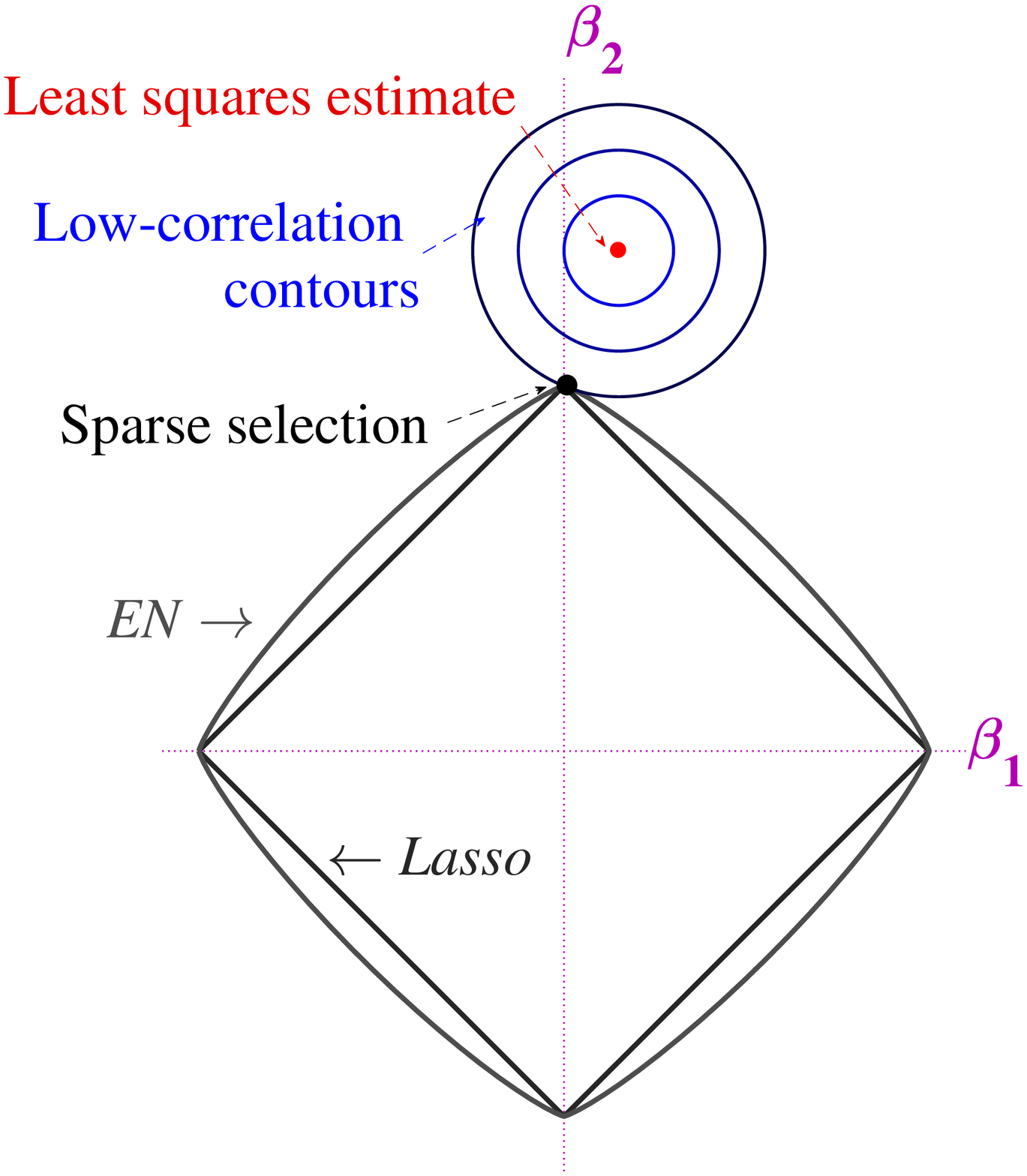}{0.44\columnwidth}{(a)}
\fig{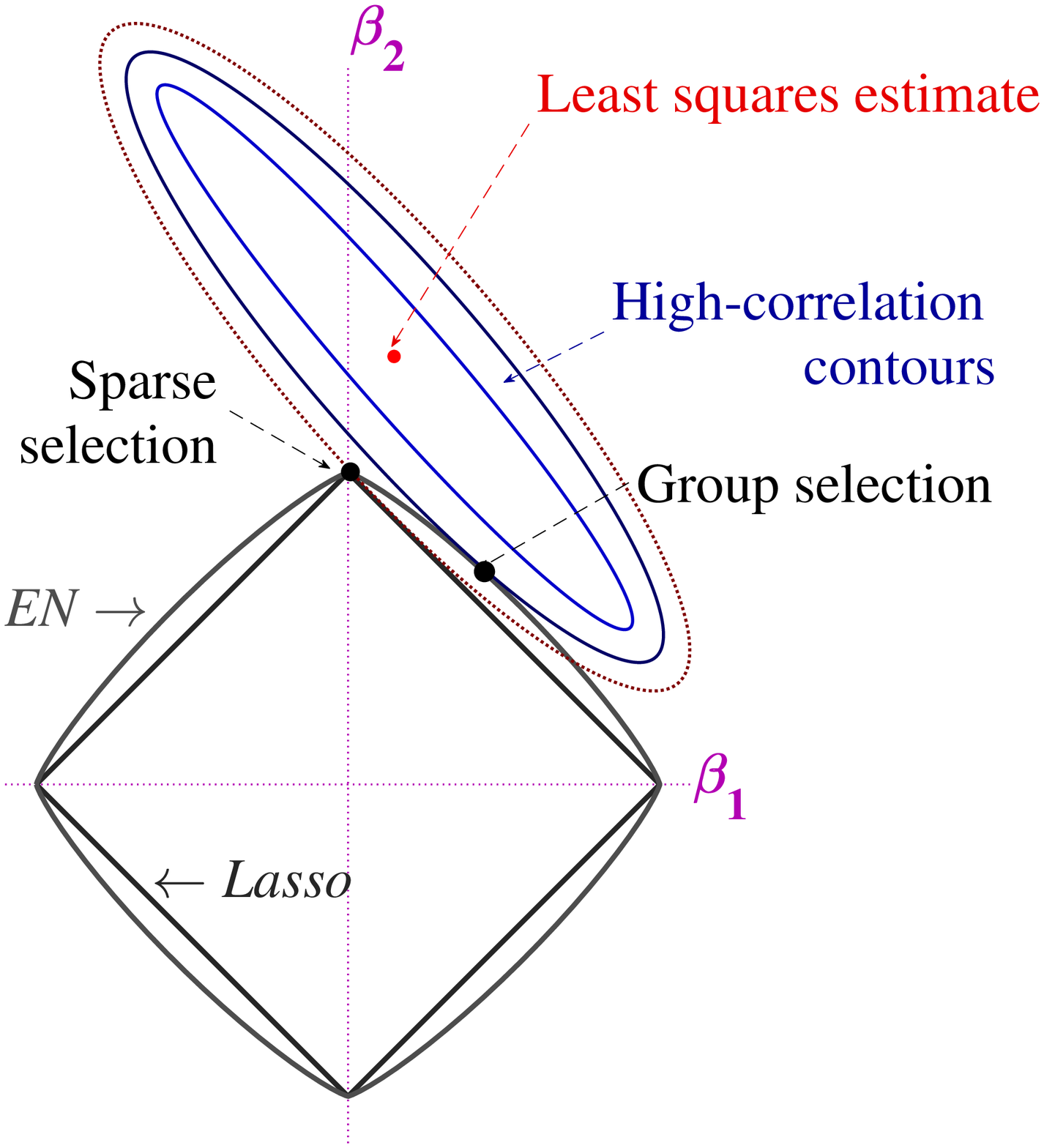}{0.48\columnwidth}{(b)}
}
\caption{(Color online) The Lasso solution is often at the vertices (corners) but the EN solution can occur on edges as well, depending on the correlations among the variables. In the uncorrelated case of (a), both the Lasso  and the EN has sparse solution. When the predictors are highly correlated as in (b), the EN has a group selection in contrast to the Lasso.} \label{fig:selection}
\end{figure}

Recall that EN tuning parameter $\al\in[0, 1]$ offers a blend amongst  the Lasso and the ridge regression.  
The benefit of EN is its ability to select correlated variables as a group, which is illustrated in Figure~\ref{fig:selection}. The EN penalty has singularities at the vertexes like Lasso, which is a necessary property for sparse estimation. It also has strictly convex edges which then help in selecting variables as a group, which is a useful property when high correlations exists between predictors.  Moreover, for $\w = \bo 1$  (a vector of ones) and $\al=1$, \eqref{eq:pwen} results in a Lasso solution and for $\w = \bo 1$  and $\al=0$ we obtain the ridge regression \cite{hoerl1970ridge} estimator.

Then WEN solution can be computed using any algorithm that can find the non-weighted ($\w = \bo 1$) EN solution. To see this, let us write 
 $\Xw  = \X  \mathrm{diag}(\w)^{-1}$ and $ \tilde \beb = \beb  \otimes \w$. 
Then,  WEN solution is found by applying the following steps. 
\begin{enumerate} 

\item Solve the (non-weighted) EN solution on transformed  data $(\y, \Xw)$: 
\[
\tilde \beb(\lam,\al)= \underset{\beb \in \C^\pdim}{\arg \min}  \, \frac  1 2  \|\y - \Xw \beb \|_2^2   + \lam P_\al\bigl(\beb \bigr),
\]  
where $ P_\al (\beb ) = P_\al (\beb ; \bo 1_p )$ is the EN penalty. 
\item WEN solution for the original data $(\y,\X)$ is 
\[ 
\hat \beb(\lam, \al) =  \tilde \beb \oslash \w
\]
\end{enumerate}

Yet, the standard (i.e., non-weighted, $w_j \equiv 1$ for $j=1,\dots, p$)  EN estimator 
may perform inconsistent variable selection. The EN solution  depends  largely on $\lam$ (and $\al$) and tuning these parameters optimally is a difficult problem. 
The adaptive Lasso  \cite{zou2006ALasso}  obtains oracle variable selection property by using cleverly chosen adaptive weights for regression coefficients in the $\ell_1$-penalty.  We extend this idea to WEN-penalty, coupling it with {\sl active set approach}, where WEN is applied to only nonzero (active) coefficients.   
Our proposed adaptive EN uses data dependent weights, defined as 
\beq \label{eq:weights}
\hat w_j =  \begin{cases} 1/| \hat \beta_{\mathrm{init},j}| ,  & \mbox{ $\hat \beta_{\mathrm{init},j} \neq 0$} \\  
\infty, & \mbox{ $\hat  \beta_{\mathrm{init},j} = 0$}  \end{cases} , \ j =1, \ldots, \pdim
\eeq 
Above $\bebh_{\mathrm{init}} \in \mathbb{C}^\pdim$ denotes a sparse initial estimator of $\beb$. The idea is that only nonzero coefficients are exploited, i.e., basis vectors 
with $\hat \beta_{\mathrm{init},j} = 0$ are omitted from the model, and thus the dimensionality of the linear model is reduced from $\pdim$ to $\kdim = \| \bebh_{\mathrm{init}} \|_0$. 
Moreover, larger weight means that the corresponding variable is penalized more heavily.  The vector $\hat \w = (\hat w_1,\ldots, \hat w_\pdim)^\top$ can be written compactly as 
\beq \label{eq:weights2}
\hat \w =  \bo 1_\pdim \oslash  \big|  \hat  \beb_{\mathrm{init}} \big | ,
\eeq
where notation $| \beb  | $ means element-wise application of the absolute value operator on the vector, 
i.e.,  $ |\beb| = (| \be_1|, \ldots, |\be_\pdim |)^\top$.

\section{Complex-valued LARS method for weighted Lasso} \label{sec:clarsen}

In this section, we  develop the c-LARS-WLasso algorithm which is a complex-valued extension of the LARS algorithm \cite{efron2004LAR} for weighted Lasso framework. This is then used to construct a  complex-valued pathwise (c-PW-)WEN algorithm. These methods  compute solution \eqref{eq:pwen} at particular penalty parameter $\lam$ values, called {\sl knots}, at which a new variable enters (or leaves) the active set of nonzero coefficients.  Our c-PW-WEN exploits the c-LARS-WLasso as its core computational engine.

Let  $\hat \beb(\lam)$ denote a solution to \eqref{eq:pwen} for some fixed value $\lam$ of the penalty parameter in  the case that  Lasso penalty is used ($\al=1$) with unit weights (i.e., $\w = \bo 1_\pdim$).  Also recall that  predictors are normalized so that $\| \x_j \|^2_2 =1$, $j=1,\ldots, p$. Then note that the solution $\hat \beb(\lam)$  needs to verify the generalized Karush-Kuhn-Tucker conditions. That is, $\hat \beb(\lam)$ is a solution to \eqref{eq:pwen} if and only if  it verifies the zero sub-gradient equations  given by
\beq 
\langle \x_{j} , \res(\lam) \rangle  =  \lam \hat s_j  \quad \mbox{for } j = 1,\ldots,\pdim \label{eq:LSestim1}
\eeq 
where $\res(\lam) =  \y-\X \hat \beb(\lam) $ and $\hat s_j \in \ssgn \{ \hat \beta_j(\lam)\}$, meaning that $\hat s_j =e^{\im \theta}$, where $\theta= \mathrm{arg}\{\hat \beta_j(\lam)\}$, if $\hat \beta_j(\lam) \neq 0$ and some number  inside the unit circle, 
$\hat s_j  \in \{ x \in \C : |x | \leq 1\}$, otherwise. Taking absolute values of both sides of equation \eqref{eq:LSestim1}, one notices that at the solution, condition $ |\langle \x_{j} , \res(\lam) \rangle  | = \lam$ holds for the active predictors whereas  $ |\langle \x_{j} , \res(\lam) \rangle  | \leq \lam$ holds for non-active predictors. Thus as $\lam$ decreases and more predictors are joined to the active set, the set of active predictors  become less correlated with the residual. Moreover, the absolute value of the correlation, or equivalently the angle between any active predictor and the residual is the same. 
In the real-valued case, the LARS method  exploits this feature and linearity of the Lasso path to compute the  knot-values, i.e., the value of the penalty parameters where there is a change in the active set of predictors.

\begin{figure}[hbt]
	\centering
	\includegraphics[width=0.97\columnwidth]{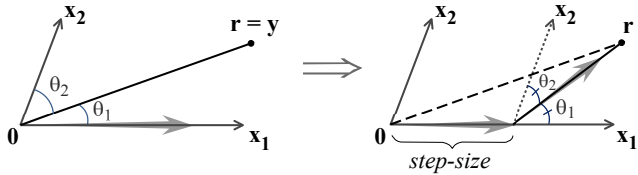}
	\caption{Starting from all zeros, LARS picks predictor $\x_1$ that makes least angle (i.e., $\theta_1 < \theta_2$) with residual $\res$ and moves in its direction until $\theta_1 = \theta_2$ where LARS picks $\x_2$ and changes direction. Then, LARS repeats this procedure and selects next equicorrelated predictor and so on until some stopping criterion.}
	\label{fig:larsidea}
\end{figure}

Let us briefly recall the main principle of the LARS algorithm. LARS starts with a model having no variables (so $\beb=\bo 0$) and picks a predictor that has maximal correlation (i.e., having smallest angle) with the residual $\res=\bo y$. Suppose the predictor $\x_1$ is chosen. Then, the magnitude of the coefficient of the selected predictor is increased (toward
its least-squares value)  
until one has reached a  step size such that another predictor (say, predictor  $\x_2$) has the  same absolute value of correlation with the evolving residual 
$\res = \y - \mathrm{(stepsize)} \, \x_1$,
 i.e., the updated residual makes equal angles with both predictors as shown in Fig. \ref{fig:larsidea}. Thereafter, LARS moves in the new direction, which keeps the evolving residual equally correlated (i.e., equiangular) with selected predictors, until another predictor becomes equally correlated with the residual. After that, one can repeat this process until all predictors are in the model or to specified sparsity level.

We first consider the {\sl weighted Lasso (WLasso)} problem (so $\al=1$) by letting $\X  = \X  \mathrm{diag}(\w)^{-1}$ and $ \beb = \beb  \otimes \w$.  
We write $\hat \beb(\lam)$ for the solution of the optimization problem \eqref{eq:pwen} in this case. Let $\lam_0$ denotes the smallest value of $\lam$ such that all coefficients of the WLasso solution are zero, i.e., $\hat \beb(\lam_0)=\bo 0$.  It is easy to see that $\lam_0= \max_j | \langle \x_j, \y \rangle | $ for $j=1,2,\dots,\pdim$ \cite{hastie2015stat}.
Let $\setA = \supp\{\hat \beb(\lam)\}$ denote the {\paino active set} at the regularization parameter value $\lam < \lam_0$. 
The {\paino knots}  $\lam_1> \lam_2 > \cdots > \lam_K$ are defined as smallest values of the penalty parameters after which there is a change in the set of active predictors, i.e.,  the order of sparsity changes.   The active set at a knot $\lam_k$ is denoted by $\setA_{k} =  \supp\{\hat \beb(\lam_k)\}$.   
The active set $\setA_1$ thus contains a single index as $\setA_1 = \{ j_1 \}$, where $j_1$ is predictor that becomes active first, 
i.e., 
\[
j_1 = \underset{j \in \{1, \ldots, p\}}{\arg \max} | \langle \x_j , \y \rangle | . 
\]
 By definition of the knots, one has that  $\setA_{k} = \supp\{\hat \beb(\lam_k)\}$ $\forall \lam \in (\lam_{k-1},\lam_k]$ and    $\setA_{k} \neq \setA_{k+1}$  for all $k=1,\ldots,K$.

The c-LARS-WLasso outlined in Algorithm~\ref{algo:LAR} is a straightforward generalization of LARS-Lasso algorithm to complex-valued and weighted case. 
It does not have the same theoretical guarantees as its real-valued counterpart to solve the exact values of the knots. Namely, LARS algorithm uses the property that (in the real-valued case) the Lasso regularization path is continuous and piecewise linear with respect to $\lam$; see \cite{efron2004LAR, rosset2007, Ryantibshirani2011, tibshirani2013lasso}. 
In the complex-valued  case,  however, the solution path between the knots is not necessarily linear \cite{panahi2012fast}. 
Hence the c-LARS-WLasso may not give precise values of the knots in all the cases. However, simulations validate that the algorithm finds the knots with reasonable precision. Future work is needed to provide theoretical guarantees of the algorithm to find the knots.

\setlength{\textfloatsep}{11pt}
\begin{algorithm}[ht]
    \SetAlgoHangIndent{0pt}
	\caption{c-LARS-WLasso algorithm} \label{algo:LAR}
	\SetKwInOut{Input}{input}\SetKwInOut{Output}{output}\SetKwInOut{Init}{initialize } 
	\Input{$\y\in\C^{n}$, $\X\in\C^{n\times p}$, $\w \in\R^{\pdim}$ and $\kdim$.}
	\Output{$\{ \setA_k, \lam_k, \bebh(\lam_k) \}_{k=0}^K$}

		\Init{$\beb^{(0)} = \bo 0_{\pdim \times 1}$, $\setA_0 = \{ \emptyset \}$, $\Del = \bo 0_{\pdim \times 1}$, the residual $\res_{0}= \y$. Set $\X \gets \X  \mathrm{diag}(\w)^{-1}$.}

	Compute  $ \lam_0  =\max_j  |\langle \x_j,\res_{0}\rangle|$  and  $j_1 = \arg \max_j  |\langle \x_j,\res_{0}\rangle|$, where $j = 1,\ldots,p$.
	\For{$k=1,\ldots,K$}{
		Find the active set $\setA_k = \setA_{k-1} \cup \{j_k\}$ and its least- squares direction $\delb$ to have $[\Del]_{\setA_{k}} = \delb$:
		\[
		\delb = \frac{1}{\lam_{k-1}}  (\Xa{k}^\hop \Xa{k})^{-1} \Xa{k}^\hop \res_{k-1}. 
		\] 

		Define vector  $ \beb (\lam) = \beb^{(k-1)}  + (\lam_{k-1}-\lam) \Del,$ for $0 < \lam \leq \lam_{k-1}, $
		and the corresponding residual as 
		\begin{align*} 
		\res(\lam) &=  \y - \X \beb(\lam) \\  
		&= \y - \X  \beb^{(k-1)}  -  (\lam_{k-1}-\lam) \X \Del  \\ &= \res_{k-1} -  (\lam_{k-1}-\lam) \Xa{k} \delb 
		\end{align*} 
        
		The knot $\lam_k$ is the largest $\lam$-value
		$0 < \lam \leq \lam_{k-1}$ s.t.
		\beq \label{eq:target}
	     \langle \x_\ell, \res(\lam) \rangle = \lam e^{\im \theta},    \quad \ell  \not \in \setA_k 
		\eeq 
		where a new predictor (at index  $j_{k+1} \not \in \setA_k $) becomes active, thus verifying 
		$   | \langle \x_{j_{k+1}}, \res(\lam_k) \rangle | = \lam_k$ from \eqref{eq:target}.

		Update the values at  a knot $ \lam_k$:
		\begin{align*} 
		\beb^{(k)}  &= \beb^{(k-1)} + (\lam_{k-1} - \lam_k)\Del \\ 
		\res_{k} &= \y - \X \beb^{(k)}. 
		\end{align*}
		The Lasso solution is $\bebh(\lam_k) = \beb^{(k)} $. 
	}
	
	$\{ \bebh(\lam_k) = \bebh(\lam_k) \oslash \w \}_{k=0}^K $
	
\end{algorithm}

Below we discuss how to solve the knot $\lam_k$ and the index $j_{k+1}$ in {\sl step~5} of the c-LARS-WLasso algorithm. 

{\it Solving step 5:}  First we note that 
\begin{align}  \label{eq:apu} 
\langle \x_\ell, \res(\lam) \rangle  &= \langle \x_\ell,\res_{k-1} -  (\lam_{k-1}-\lam)\Xa{k} \delb  \rangle \notag  \\ 
&=\langle \x_\ell, \res_{k-1}  \rangle  - (\lam_{k-1}-\lam) \langle \x_\ell, \Xa{k} \delb \rangle \notag \\
&=c_\ell -  (\lam_{k-1}-\lam) b_\ell,
\end{align} 
where we have written $c_\ell = \langle \x_\ell, \res_{k-1}  \rangle$ and $b_\ell =  \langle \x_\ell, \Xa{k} \delb \rangle$.  
First we need to find $\lambda$ for each $\ell \not \in \setA_k$, such that  $| \langle \x_\ell, \res(\lam) \rangle | = \lam$  holds. 
Due to \eqref{eq:target} and \eqref{eq:apu} this means finding $0<\lam \leq \lam_{k-1}$ such that 
\beq \label{eq:apu2}
c_\ell -  (\lam_{k-1}-\lam) b_\ell  =  \lam e^{\im \theta}. 
\eeq 
Let us reparametrize such that  $\lam=\lam_{k-1} - \gamma_\ell$. Then identifying $ 0 < \lam \leq \lam_{k-1}$ is equivalent to identifying the auxiliary variable $\gamma_\ell\geq 0$.  
Now \eqref{eq:apu2} becomes
\begin{gather*}
c_\ell -  (\lam_{k-1}-\lam_{k-1} + \gamma_\ell ) b_\ell  = (\lam_{k-1} - \gamma_\ell) \: e^{\im \theta} \\
\Leftrightarrow c_\ell - \gamma_{\ell} b_\ell = (\lam_{k-1} - \gamma_\ell) \: e^{\im \theta} \\ 
\Leftrightarrow |c_\ell - \gamma_{\ell} b_\ell|^2 = (\lam_{k-1} - \gamma_\ell)^2 \\  
\Leftrightarrow |c_\ell|^2 - 2 \gamma_{\ell} \mathrm{Re}(c_\ell b_\ell^*) + \gamma_\ell^2 |b_\ell|^2 = \lam_{k-1}^2 - 2 \lam_{k-1} \gamma_\ell  + \gamma_\ell^2.  
\end{gather*} 
The last equation implies that $\gamma_{\ell}$ can be found by solving the roots of the second-order polynomial equation,  $A \gamma_\ell^2 + B \gamma_\ell + C = 0$,  
where $A = |b_\ell|^2 - 1, \: B = 2 \lam_{k-1} - 2 \mathrm{Re}(c_\ell b_\ell^*)$ and $C = |c_\ell|^2 - \lam_{k-1}^2$. 
The roots are $\gamma_\ell = \{\gamma_{\ell1}, \gamma_{\ell2}\}$ and the final value of $\gamma_\ell$ will be 
\[
\gamma_{\ell} = \begin{cases}  \min(\gamma_{\ell1}, \gamma_{\ell2}), &\mbox{if $\gamma_{\ell1} > 0$ and $\gamma_{\ell2} > 0$} \\    \{\max(\gamma_{\ell1}, \gamma_{\ell2})\}_+, &\mbox{otherwise} \end{cases} 
\]
where $(t)_+= \max(0,t)$ for $t \in \R$. Thus finding largest $\lambda$ that verifies \eqref{eq:target} is equivalent to finding smallest non-negative $\gamma_\ell$.  Hence the variable that enters to  active set $\setA_{k+1}$ is    $j_{k+1} = \arg \min_{\ell \not \in \setA_k} \gamma_\ell$ and the knot is thus  $\lam_k=\lam_{k-1} - \gamma_{j_{k+1}}$.
Thereafter, solution $\bebh(\lam_k)$ at the knot $\lam_k$ is simple to find in {\sl step~6}.

\section{Complex-valued Pathwise Weighted Elastic Net} \label{sec:pwen}

Next we develop a complex-valued and weighted version of PW-LARS-EN algorithm proposed  in \cite{tabassum2017pathwise}, and 
referred to as  c-PW-WEN algorithm. 
Generalization to complex-valued case is  straightforward.  The essential difference is that  c-LARS-WLasso  Algorithm~\ref{algo:LAR} is used instead of the (real-valued) LARS-Lasso algorithm. The algorithm finds the $K^{th}$ knot $\lam_\kdim$ and the corresponding WEN solutions  at a dense grid of EN tuning parameter values $\al$ and then picks final solution for best $\al$-value.

Let $\lam_0(\al)$ denotes the smallest value of $\lam$ such that all coefficients in the WEN estimate are zero, i.e., $ \bebh(\lam_0, \al)= \bo 0$. The value of $\lam_0(\al)$ can be expressed in closed-form \cite{hastie2015stat}:
\[
\lam_0(\al) = \max_j \frac{1}{\al} \bigg| \frac{\langle \x_j, \y \rangle }{w_j} \bigg|, \qquad j=1,\ldots,\pdim .
\]
The c-PW-WEN algorithm computes $\kdim$-sparse WEN solutions for  a set of $\al$ values in a dense grid 
\beq \label{eq:al_grid}
[\al] = \{ \al_{i} \in [1, 0) \ :  \   \al_{1}=1 <  \cdots < \al_{\mdim} <0  \}.
\eeq 
Let $\setA(\lam) = \supp\{\hat \beb(\lam, \al)\}$ denote the {\paino active set} (i.e., nonzero elements of WEN solution) for a given fixed regularization parameter value $\lam \equiv \lam(\al) < \lam_0(\al)$ and for given $\al$ value in the grid $[\al]$. 
The knots  $\lam_1(\al) > \lam_2(\al) > \cdots > \lam_\kdim(\al)$ are the border values of the regularization parameter after which there is a change in the set of active predictors. 
Since $\al$ is fixed we drop the dependency of the penalty parameter on $\al$ and simply write $\lambda$ or $\lam_\kdim$ instead of $\lambda(\al)$ or $\lambda_\kdim(\al)$. The reader should however keep in mind that the value of the knots are different for any given $\al$. 
The active set at a knot $\lam_k$ is then denoted shortly by $\setA_{k} \equiv \setA(\lam_k)=  \supp\{\hat \beb(\lam_k,\al)\}$.    
Note that $\setA_{k} \neq \setA_{k+1}$ for all $k=1,\ldots,\kdim$. Note that it is assumed that a non-zero coefficient does not leave the active set for  any value $\lam > \lam_\kdim$, that is, the sparsity level is increasing from $0$ (at $\lam_0$) to $K$ (at $\lam_K$). 

First we let that Algorithm~\ref{algo:LAR} can be written as $\text{c-LARS-WLasso}( \y,\X, \w, \kdim)$ then we can write it as let  
\[ 
\{\lam_k, \bebh(\lam_k) \} = \text{c-LARS-WLasso}( \y,\X, \w) \big|_k  
\] 
for extracting the $k^{th}$ knot (and the corresponding solution) from a  sequence of the  knot-solution pairs found by the c-LARS-WLasso algorithm. 
Next note that we can write the EN objective function in  {\paino augmented form} 
as follows:   
\beq\label{eq:augEN} 
\frac 1 2 \| \y - \X \be \|_2^2 + \lam P_\al(\be) = 
 \ \frac  1 2 \| \y_a   - \X_a(\eta)  \be \:\|_2^2   + \gam \big\| \be \big\|_1 
\eeq 
where  
\beq \label{eq:gam_eta} 
	\gam  = \lam \al \qquad \text{and} \qquad \eta = \lam (1-\al), 
\eeq 
are new parameterizations of the tuning and shrinkage parameter pair $(\al,\lam)$,  and  
\[ 
\y_a = \bmat \y \\ \bo 0 \emat \qquad \text{and} \qquad  \X_a(\eta)  =  \bmat \X \\ \sqrt{\eta} \, \bo I_\pdim  \emat 
 \] 
are  the {\paino augmented}   forms of the response vector $\y$ and the predictor matrix $\X$, respectively. Note that \eqref{eq:augEN} resembles the Lasso objective function with $\y_a \in \R^{\ndim + \pdim}$ and that
 $\X_a(\eta)$ is an $(\ndim + \pdim) \times \pdim$ matrix.

It means that we can compute the $\kdim$-sparse WEN solution at $\kdim^{th}$ knot for fixed $\al$ using the c-LARS-WLasso algorithm. 
Our c-PW-WEN  method is given in \autoref{algo:pwen}. It computes the WEN solutions at the knots over a dense grid  \eqref{eq:al_grid} of $\al$  values. 
After Step 6 of the algorithm we have solution at the $\kdim^{th}$ knot for  a given $\al_i$ value on the grid. 
Having  the solution $\bebh(\lam_\kdim,\al_i)$ available, we then in steps 7 to 9, 
compute the  residual sum of squares (RSS) of the debiased WEN solution at the $\kdim^{th}$ knot (having $\kdim$ nonzeros):  
\[
\textsc{RSS}(\al_i) = \|\y - \X_{\setA_\kdim} \bebh_{\texttt{LS}}(\lam_{\kdim},\al_i) \|^2_2,
\]
where $\setA_\kdim=\supp( \hat \beb(\lam_\kdim,\al_i))$ is the active set at the $\kdim^{th}$ knot and $\bebh_{\texttt{LS}}(\lam_{\kdim},\al_i)$ is 
the {\paino debiased LSE}, defined as 
\beq \label{eq:lse} 
\bebh_{\texttt{LS}}(\lam_{\kdim},\al_i) = \X_{\setA_\kdim}^{+} \y,
\eeq 
where $\X_{\setA_\kdim} \in \C^{n \times \kdim}$ consists of the $\kdim$ active columns of $\X$  associated with the active set $\setA_\kdim$ and $\X_{\setA_\kdim}^{+}$ 
denotes its  Moore-Penrose pseudo inverse.

While sweeping through the grid of $\al$ values  and computing the WEN solutions $\hat \beb(\lam_\kdim,\al_i)$, we choose our best candidate solution as  the WEN estimate $ \bebh(\lam_\kdim,\al_\imath)$  that had the smallest RSS value,
i.e., $ \imath = \ \arg \min_i  \textsc{RSS}(\al_i)$, where  $i \in \{ 1,\ldots,\mdim\}$.

\begin{algorithm}[hbt]
	\DontPrintSemicolon
	\caption{c-PW-WEN algorithm.}\label{algo:pwen}
	\SetKwInOut{Input}{input}\SetKwInOut{Output}{output}
	\BlankLine                                                   
	\Input{$\quad \y\in\C^{n}$, $\X\in\C^{n\times p}$, $\w \in\R^{\pdim}$, $[\al] \in\R^{\mdim}$ (recall $\al_1=1$), $\kdim$ and $\textsc{debias}$. 
	}
	\BlankLine
	\Output{\quad $\bebh_\kdim \in\C^{\pdim}$ and $\setA_\kdim  \in\R^{\kdim}$.}
	
\BlankLine
$ \{ \lam_k(\al_1), \bebh(\lam_k,\al_1) \}_{k=0}^K = \text{c-LARS-WLasso}\big( \y,\X, \w, \kdim \big) $

\BlankLine
   \For{$i=2$ \KwTo $\mdim$}{
      \For{$k=1$ \KwTo $K$}{

\BlankLine
 $ \tilde \eta_k  =  \lam_k(\al_{i-1}) \cdot (1-\al_i) $ 
  \BlankLine
$\big\{ \gam_k , \bebh(\lam_k,\al_i) \big\} = \text{c-LARS-WLasso}\big( \y_a, \X_a(\tilde \eta_k), \w ) \big|_k $
\BlankLine
$\lam_k(\al_{i}) = \gam_k/\al_i \quad$
}
\BlankLine
		$\setA_\kdim  = \supp\{\hat \beb(\lam_\kdim, \al_i) \}$
			
		$\bebh_{\texttt{LS}}(\lam_{\kdim},\al_i) = \X_{\setA_\kdim}^{+} \y$
		
		$\textsc{RSS}(\al_i) = \|\y - \X_{\setA_\kdim} \bebh_{\texttt{LS}}(\lam_{\kdim},\al_i) \|^2_2,$

}

			\BlankLine
	$\imath = \ \arg \min_i  \textsc{RSS}(\al_i)$
	
	$\bebh_\kdim = \bebh(\lam_\kdim,\al_\imath) \qquad$ and 
	$\qquad \setA_\kdim = \supp\{\bebh_\kdim\}$
	
  \lIf{$\textsc{debias}$}
{
		$\bebh_{\setA_\kdim} = \X_{\setA_\kdim}^{+} \y$
}

\end{algorithm}

\section{Sequentially Adaptive Elastic Net}  \label{sec:saen}

Next we turn our attention on how to choose the adaptive (i.e., data dependent) weights in c-PW-WEN. 
In adaptive Lasso \cite{zou2006ALasso}, one ideally uses the LSE or, if $\pdim > \ndim$, the Lasso as an initial estimator $\bebh_{\mathrm{init}}$ to construct the weights given in \eqref{eq:weights}.  
The problem is that both the  LSE and the Lasso estimator have very poor accuracy (high variance) when there exists high correlations between predictors,  which is the condition 
we are concerned in this paper. This lowers the probability of exact recovery of the adaptive Lasso significantly.

To overcome the problem above, we devise  a sequential adaptive elastic net  (SAEN) algorithm that obtains the $\kdim$-sparse solution in a sequential manner decreasing the sparsity level of the solution at each iteration and using the previous solution as adaptive weights for c-PW-WEN.   The SAEN  is described in \autoref{algo:saen}. SAEN runs the c-PW-WEN algorithm three times. At first step,  it  finds a standard (unit weight) c-PW-WEN solution for $3\kdim$ nonzero (active) coefficients which we refer to as initial EN solution $\beTKi$. The obtained solution determines the adaptive weights via \eqref{eq:weights} (and hence the active set of $3\kdim$ nonzero coefficients) which is used in the second  step to computes the c-PW-WEN solution that has 
$2\kdim$ nonzero coefficients. This again determines the adaptive weights via \eqref{eq:weights} (and hence the active set of $2\kdim$ nonzero coefficients) which is used in the second  step to compute the c-PW-WEN solution that has the desired $\kdim$ nonzero coefficients. 
It is important to notice that since we start from a solution with $3\kdim$ nonzeros, it is quite likely that the true $\kdim$ non-zero  coefficients will be included in the active set of 
$\beTKi$ which is computed in the first step of the SAEN algorithm. 
\added{Note that the choice of $3 K$ is similar to CoSaMP algorithm \cite{Needell2009CoSaMP} which also uses $3K$ as an initial support size.  Using  $3K$ also usually guarantees that  $\mathbf{X}_{\mathcal A_{3K}}$ is  well conditioned which may not be the case if larger value than $3K$ is chosen.}

\begin{algorithm}[hbt]
     \SetAlgoHangIndent{0pt}
\caption{SAEN algorithm}\label{algo:saen}
\SetKwInOut{Input}{input}\SetKwInOut{Output}{output }
\BlankLine
\Input{$\quad \y\in\C^{n}$, $\X\in\C^{n\times p}$, $[\al] \in\R^{\mdim}$ and $\kdim$.} 
\BlankLine
\Output{\quad $\bebh_\kdim \in\C^{p}$}
\BlankLine 
$\big\{ \beTKi, \setA_{3 \kdim} \big\}  = \text{c-PW-WEN}\big( \y, \X, \bo 1_\pdim, [\al], 3 \kdim, 0 \big)$ 

$\big\{ \bebh, \setA_{2 \kdim} \big\}  = \text{c-PW-WEN}\big( \y, \X_{\setA_{3 \kdim}},  \bo 1_{3 \kdim} \oslash    \big| \beTKi_{\setA_{3 \kdim}} \big|  , [\al], 2 \kdim, 0 \big)$ 

$\big\{ \bebh_{\kdim}, \setA_{\kdim} \big\}  = \text{c-PW-WEN}\big( \y, \X_{\setA_{2 \kdim}},  \bo 1_{2 \kdim} \oslash \big |\bebh_{\setA_{2 \kdim}} \big|, [\al], \kdim, 1 \big)$ 

\end{algorithm}

\section{Single-snapshot compressive beamforming}   \label{sec:cbf}

Estimating the source location, in terms of its DoA, plays an important role in many applications. 
In \cite{malioutov2005sparse},  it was observed that CS algorithms can be applied for DoA estimation (e.g. of sound sources) using sensor arrays when the array output $\y$ is be expressed as sparse (underdetermined) linear model by discretizing the DoA parameter space. This approach is referred to as compressive beamforming (CBF), and it has been subsequently used in a series of papers (e.g.,  \cite{fortunati2014single, Gerstoft2014Compressive, ollila:2015,ollila:2015b,Gerstoft2015Multiple, Seong2016spherical, Gerstoft2017Adaptive, Gerstoft2017Coherent}). 

In CBF, after finding the sparse regression estimator, its support can be mapped to the DoA estimates on the grid. Thus the DoA estimates in CBF are always selected from the resulting finite set of discretized DoA parameters. Hence the resolution of CBF is dependent on the density of the grid (spacing $\delTh$ or grid size $p$). Denser grid implies large mutual coherence of the basis vectors $\x_j$ (here equal to the steering vectors for the DoAs on the grid) and thus a poor recovery region for most sparse regression techniques.  

The proposed SAEN estimator can effectively mitigate the effect of high mutual coherence caused by discretization of the DoA space with significantly better performance than state-of-the-art compressed sensing algorithms. This is illustrated in Section~\ref{sec:res} via extensive simulation studies using  challenging multi-source set-ups of closely-spaced sources and large variation of source powers.

We assume narrowband processing and a far-field source wave impinging on an array of sensors with known configuration.
The sources are assumed to be located in the far-field of the sensor array (i.e., propagation radius $\gg$ array size). 
A uniform linear array (ULA) of $\ndim$ sensors (e.g. hydrophones or microphones) is used for estimating the DoA $\theta  \in [-90^{\circ},90^{\circ})$ of the source with respect to the array axis.
The array response (steering or wavefront vector) of ULA for a source from DoA (in radians) $\theta \in [-\pi/2, \pi/2) $ is given by
\[
\a\bigl(\theta\bigr) \triangleq \cfrac{1}{\sqrt{n}}\: \bigl[1,e^{ \im \pi\sin\theta},\ldots,e^{ \im \pi(n-1)\sin\theta}\bigr]^\top,  
\]
where we assume half a wavelength inter-element spacing between sensors. 
We consider the case that $\kdim < \ndim$ sources from distinct DoAs $\thb = (\theta_1,\ldots,\theta_\kdim)$ arrive at a sensor at some time instant $t$. 
A single-snapshot obtained by ULA can then be modeled as \cite{Mecklen1997coherence} 
\beq \label{eq:snapshot} 
\y(t) = \A(\thb) \s(t) + \vepsb(t)
\eeq
where $\thb=(\theta_1,\ldots,\theta_\kdim)^\top$ collects the DoAs, the matrix $\A(\thb) =  [ \a(\theta_1) \cdots \a(\theta_\kdim) ] \triangleq \bo A \in \C^{n \times \kdim}$ is the dictionary of replicas also known as the array steering matrix, $\s(t) \in \C^\kdim$ contains the  source waveforms and $\vepsb(t)$ is complex noise at time instant $t$. 

Consider an angular grid of size $\pdim$ (commonly $\pdim \gg \ndim$) 
of look directions of interest: 
\[
[\vartheta] = \{ \vartheta_{i} \in [-\pi/2, \pi/2) \ :  \   \vartheta_{1} <  \cdots < \vartheta_{\pdim} \}. 
\]
Let the $i^{th}$ column of  the measurement matrix $\X$  in the model \eqref{eq:linear} be the array response for look direction $\vartheta_{i}$, so $\x_i = \a(\vartheta_{i})$. Then, if the true source DoAs are contained in the angular grid, i.e.,  $ \theta_i \in [\vartheta]$ for $i=1,\ldots,\kdim$, then the snapshot $\y$ in  \eqref{eq:snapshot} (where we drop the time index $t$) can be equivalently modeled by \eqref{eq:linear} as
\[
\y = \X \beb + \vepsb
\]
where $\beb$ is exactly $\kdim$-sparse ($\| \beb \|_0=\kdim$) and  nonzeros elements of $\beb$  maintain  the source waveforms $\s$.  
Thus, identifying the true DoAs is equivalent to identifying the nonzero elements of $\beb$, which we refer to as {\paino CBF-principle}. Hence, sparse regression and CS methods can be utilised for estimating  the DoAs based on a single snapshot only.  We assume that the number of sources $\kdim$ is known a priori.

Besides SNR, also the  size $\pdim$ or spacing $\Delta \theta$ of the grid greatly affect the performance of CBF methods. 
The cross-correlation (coherence) between the true steering vector and steering vectors on the grid depends on both  the grid spacing and obliqueness of the target DoAs w.r.t. to the array. Moreover, the values of cross-correlations in the gram matrix $| \X^\hop \X |$ also depend on the distance between array elements and configuration of the sensors array \cite{Gerstoft2014Compressive}.
Let us construct a measure, called {\sl maximal basis coherence (MBC)},  defined as  the maximum absolute value of the cross-correlations among the true steering vectors $\a(\theta_j)$    and the basis  $\a(\vartheta_{i})$, $\vartheta_i \in [\vartheta]\setminus \{ \theta_j\} $, $j \in \{1, \ldots, K\}$, 
\beq \label{eq:MBC} 
\mathrm{MBC} =  \max_{j } \max_{ \vartheta \in  [\vartheta]  \setminus \{ \theta_j \} }  \,  \big|\a(\theta_j)^\hop \,\a(\vartheta) \big|. 
\eeq 
Note that steering vectors $\a(\theta)$, $\theta \in [-\pi/2, \pi/2)$,  are assumed to be normalized  ($\a(\theta)^\hop \,\a(\theta) = 1$). MBC value measures the obliqueness of 
the incoming DoA to the array. The higher the MBC value, the more difficult it is for any CBF method to distinguish the true DoA  in the grid.  Note that the value of MBC also depends on the grid spacing 
$\Delta \theta$.

\begin{figure}[hbt]
	\centering
	\includegraphics[width=0.98\columnwidth]{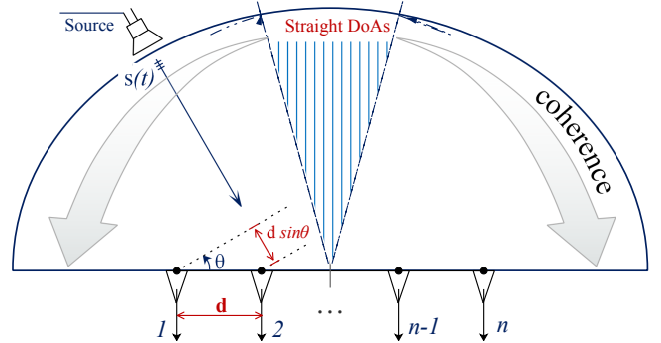}
	\caption{(Color online) The straight and oblique DoAs exhibit different basis coherence.}
	\label{fig:DoAnorVob}
\end{figure}

Fig. \ref{fig:DoAnorVob} shows the geometry of DoA estimation problem, where the target DoA's have varying basis coherence and it increases with the level of obliqueness (inclination) on either side of the normal to the ULA axis. We define a {\sl straight DoA} when the angle of incidence of the  target DoAs is in the shaded sector in Fig.~\ref{fig:DoAnorVob}. This the region where 
the MBC has lower values.  In contrast, an {\sl oblique DoA} is defined  when the angle of incidence of the target is oblique with respect to the array axis. 

Consider the case of ULA with $\ndim=40$ elements receiving two sources at straight DoAs, $\theta_1 = -6^\circ$ and $\theta_2 = 2^\circ$, or at oblique DoAs, $\theta_1 = 44^\circ$ and $\theta_2 = 52^\circ$. The above scenarios correspond to set-up 2 and set-up 3 of Section~\ref{sec:res}; see also Table~\ref{table:scene}.  
Angular separation between the DoA's is $8^\circ$ in both the scenarios.  In Table~\ref{table:ncc} we compute the correlation between the true steering vectors  and with a neighboring 
steering vector $\bo a(\vartheta)$ in the grid.   

\begin{table}[hbt]
	\centering
	\caption{ Correlation between true steering vector at DoAs $\theta_1$ and $\theta_2$ and a steering vector at  angle $\vartheta$ on the grid  in a two source scenario set-ups with either  straight or oblique DoAs. } \label{table:ncc}
	\begin{ruledtabular}
   \begin{tabular*}{0.98\columnwidth}{@{\extracolsep{\fill}} l  c c r c @{}}
			&$\theta_1$ &$\theta_2$ &$\vartheta$ &  correlation \\ 
			\midrule
			True {\sl straight DoAs}  &$-6^\circ$  		& $2^\circ$ 	& &0.071 \\
				 			  	&$-6^\circ$ 	&  			& $-7^\circ$ 	&  0.814 \\
								&			& $2^\circ$ 	& $3^\circ$ 	&  0.812 \\
			\midrule
			True {\sl oblique DoAs}  		&$44^\circ$  	&$52^\circ$  	& 		 	&  0.069 \\
								&$44^\circ$  	&  			& $45^\circ$ 	&  0.901 \\
								& 	    		&$52^\circ$  	&$53^\circ$ 	&{\bf 0.927} \\
\end{tabular*} 
\end{ruledtabular}
\end{table}
These values validate the fact highlighted in Fig. \ref{fig:DoAnorVob}. Namely, a target with  an oblique DoA w.r.t. the array has a larger maximal correlation (coherence) with the basis steering vectors. This  makes it difficult for the sparse recovery method to identify the true steering vector $\a(\theta_i)$  from the spurious steering vector $\a(\vartheta)$ that simply has a very large correlation with  the true one. Due to this mutual coherence, it may happen that neither $\a(\theta_i)$ or $\a(\vartheta)$ are assigned a non-zero coefficient value in $\bebh$ or perhaps just one of them in random fashion.

\section{Simulation studies}  \label{sec:res}

We consider seven simulation set-ups. First  five set-ups use grid spacing $\delTh =1^{\circ}$ (leading to  $p=180$ look directions in the grid $[\vartheta]$) and the last two employ sparser grid  $\delTh=2^{\circ}$ (leading to $p=90$ look directions in the grid). 
The number of sensors $n$ in the ULA are $\ndim=40$ for set-ups 1-5 and $\ndim=30$ for set-ups 6-7. Each set-up has $\kdim \in \{2, 3, 4\}$  sources at different  (straight or oblique) DoAs $\thb=(\theta_1,\ldots,\theta_{\kdim})^\top$  and the  source waveforms are generated as $s_k =  |s_k | \cdot e^{ \im  \mathrm{Arg}(s_k)}$, where source powers $|s_k| \in (0,1]$ are fixed for each set-up but the source phases are randomly generated for each Monte-Carlo trial as $\mathrm{Arg}(s_k) \sim \mathrm{Unif}(0,2 \pi)$, for $k=1,\ldots,K$. \autoref{table:scene} specifies the values of DoA-s and power of the sources used in the set-ups. Also the MBC values \eqref{eq:MBC} are reported for each case.  

\begin{table}[hbt]
\setlength\extrarowheight{0pt} 
\centering
	\setlength{\textfloatsep}{-0.1cm}
    \caption{Details of all the set-ups tested in this paper. First five set-ups have grid spacing $\delTh=1^{\circ}$ and last two $\delTh=2^{\circ}$.} \label{table:scene}
    \begin{ruledtabular}
    {\begin{tabular*}{0.99\columnwidth}{@{\extracolsep{\fill}} c |  c  c c @{}}
     Set-ups &$|s_i|$ &$\thb$ $[^\circ]$ &MBC\\
    \midrule 
        1  &$[0.9,1,1]$ &$[-5,3,6]$ &0.814 \\
        2  &$[0.9,1]$ &$[-6,2]$ &0.814 \\ 
        3  &$[0.9,1]$ &$[44,52]$ &0.927 \\ 
        4  &$[0.8,0.7,1]$ &$[43,44,52]$ &0.927 \\ 
        5  &$[0.9,0.1,1,0.4]$ &$[-8.7,-3.8,-3.5,9.7]$ & 0.990 \\ 
    \midrule 
        6  &$[0.8,1,0.9,0.4]$ &$-[48.5,46.4,31.5,22]$ &0.991  \\ 
        7  &$[0.7,1,0.6,0.7]$ &$[6,8,14,18]$ &0.643  \\ 
\end{tabular*} }
\end{ruledtabular}
\end{table}

The error terms $\varepsilon_i$ are i.i.d. and generated from $\C \mathcal N(0,\sigma^2)$ distribution, where the noise variance $\sigma^2$ depends on the signal-to-noise ratio (SNR) level in decibel (dB), given by 
$\mathrm{SNR}$(dB) $  = 10\log_{10} (\sigma_s^2/\sigma^2)$, 
where $\sigma_s^2 = \frac{1}{\kdim} \bigl\{|s_1|^2 +|s_2|^2+\dots+|s_\kdim|^2\bigr\}$ denotes the average source power. An SNR level of 20 dB is used in this paper unless its specified otherwise and   the number of Monte-Carlo trials  is $L=1000$.

In each set-up, we evaluate the performance of all methods in recovering exactly the true support and the source powers. 
Due to high mutual coherence and due the large differences in source powers, the DoA estimation is now a challenging task. 
A key performance measure is the  (empirical) {\paino probability of exact recovery (PER)}  of all $\kdim$ sources,  defined as  
\[
\mathrm{PER} = \mathrm{ave} \{ \mathsf{I} ( \setA^\ast= \hat \setA_\kdim) \},
\]
where  $\setA^\ast$ denotes the index set of  true source DoAs on the grid and set $\hat \setA_\kdim$  the found support set, where $ |\setA^\ast| = | \hat \setA_\kdim| = \kdim$, and the average is over all Monte-Carlo trials.  
Above  $\mathsf{I}(\cdot)$ denotes the indicator function. We also compute the average  root mean squared error (RMSE) of the debiased estimate $\hat \s = \arg \min_{\s \in \mathbb{C}^\kdim} \| \y - \X_{\hat \setA_\kdim} \s \|^2 $ of the source vector as  $\mathrm{RMSE}  = \sqrt{ \mathrm{ave} \{ \| \s - \hat \s \|^2  \} }$.

\subsection{Compared methods}

This paper compares the SAEN approach to the existing well-known greedy methods, such as orthogonal matching pursuit (OMP) \cite{Tropp2007omp} and compressive sampling matching pursuit (CoSaMP) \cite{Needell2009CoSaMP}. Moreover, we also draw comparisons for two special cases of the c-PW-WEN algorithm 
to Lasso estimate that has $K$-non-zeros (i.e., $\bebh(\lam_K,1)$, computed by c-PW-WEN using $\w = \bo 1$ and $\al=1$)  
and EN estimator when cherry-picking the best $\al$ in the grid $[\al] = \{ \al_{i} \in [1, 0) \ :  \   \al_{1}=1 <  \cdots < \al_{\mdim} <0.01  \}$ (i.e.,  $\bebh(\lam_K,\al_{bst})$). 

It is  instructive to compare  the SAEN to simpler  adaptive EN (AEN) approach that simply uses adaptive weights to weight different coefficients differently in the spirit of adaptive Lasso\cite{zou2006ALasso}. This helps in understanding the effectiveness of the cleverly chosen weights and the usefulness of the three-step procedure used  by  the SAEN algorithm. 
Recall that the first step in AEN approach is compute the weights using some initial solution. 
After obtaining the (adaptive) weights the final $\kdim$-sparse AEN solution is computed.  
 We devise three AEN approaches each one using a different 
initial solution to compute the weights:

 \begin{enumerate} 
\item AEN$^{(LSE)}$ uses the weights found as
\[
\w^{(LSE)} = \bo 1 \oslash  | \X^{+} \y |,
\]
where $\X^{+}$ is Moore-Penrose pseudo inverse of $\X$. 

\item  AEN$^{(n)}$ employs weights from an initial $n$-sparse EN solution $\bebh(\lam_n, \al)$ at $n^{th}$ knot which is found by c-PW-WEN algorithm with $\w = \bo 1$. 

\item  AEN$^{(3\kdim)}$ instead uses weights calculated from an initial EN solution $\bebh(\lam_{3\kdim}, \al)$ having $3\kdim$ nonzeros as in step 1 of SAEN algorithm, but the remaining two steps of SAEN algorithm are omitted. 
\end{enumerate}

The upper bound for PER rate for SAEN is the (empirical) probability that the initial solution $\beTKi$ computed in  step~1 of  \autoref{algo:saen} 
contains the true support $\setA^\ast$, i.e., the value 
\beq \label{eq:upperbound}
\mathrm{UB} = \mathrm{ave} \big\{ \mathsf{I} \big( \setA^\ast \subset \supp( \beTKi \big) \big\}
\eeq 
where the average is over all Monte-Carlo trials. We also compute this upper bound to illustrate the ability of SAEN to pick the true $\kdim$-sparse support from the original $3\kdim$-sparse initial value. 
For  set-up~1 (cf. Table \ref{table:scene}), the average recovery results for all of the above mentioned methods are provided in Table \ref{table:sc1}.

\begin{table}[t]
	\centering
\caption{The recovery results for set-up 1. Results illustrate the effectiveness of three step SAEN approach compared to its competitors.  The SNR level was 20 dB and the upper bound \eqref{eq:upperbound} for the PER rate of SAEN is given in parentheses.} \label{table:sc1}
\begin{ruledtabular}
\begin{tabular*}{0.97\columnwidth}{@{\extracolsep{\fill}} l | c c c c c c @{}}
			 &           & SAEN  &AEN$^{(3\kdim)}$ &AEN$^{(n)}$ & \\
			\midrule
			PER    & (0.957)  & {\bf 0.864} & 0.689  &0.616 & \\
            RMSE    	&-    &{\bf 0.449}   &0.694    &0.889    & \\
			\midrule
			    &AEN$^{(LSE)}$ &EN &Lasso &OMP &CoSaMP   \\
			\midrule
			PER &0    &0.332   &0.332    &0.477  &0.140 \\
			RMSE    &1.870    &1.163    &1.163   &1.060   &35.58  \\
\end{tabular*}
\end{ruledtabular} 
\end{table}

It can be noted that the proposed SAEN outperforms all other methods and weighting schemes and recovers he true support and powers of the sources effectively. 
Note that  the SAEN's  upper bound for PER rate was 95.7\% and SAEN reached the PER rate 86.4\%. 
The results of the AEN approaches validate the need for accurate initial estimate to construct the adaptive weights.  For example, AEN$^{(3\kdim)}$ performs better 
than AEN$^{(n)}$, but much worse than the SAEN method.

\subsection{Straight and oblique DoAs}

Set-up~2 and set-up~3 correspond to the case where the targets are at straight and oblique DoAs, respectively.  Performance results of the sparse recovery algorithms are tabulated in \autoref{table:sc2a3}. As can be seen, the upper bound for PER rate of SAEN is  full 100\% percentage which means that the  true support is correctly included in $3\kdim$-sparse solution computed at Step~1 of the algorithm.  For set-up 2 (straight DoAs), all methods have almost full PER rates except CoSaMP with  67.8\%  rate.  Performance of other estimators expect of SAEN changes drastically in set-up 3 (oblique DoAs). Here  SAEN is achieving nearly perfect (${\sim}$98\%) PER rate which means reduction of 2\% compared to set-up 2.  Other methods perform poorly. For example, PER rate of Lasso drops from near 98\%  to 40\%.  Similar behavior is observed for EN, OMP and CoSaMP.

\begin{table}[hbt]
    \setlength\extrarowheight{2pt}
	\setlength{\textfloatsep}{-0.1cm}
	\centering	\caption{Recovery results for set-ups 2 - 4. Note that for oblique DoA's (set-ups  3 and 4), the SAEN method outperform the other methods and has a perfect recovery results 
	for set-up 2 (straight DoAs).  SNR level is 20 dB. The upper bound \eqref{eq:upperbound} for the PER rate of SAEN is given in parentheses.}  \label{table:sc2a3}
	\begin{ruledtabular}
    {\begin{tabular*}{0.97\columnwidth}{@{\extracolsep{\fill}} l | c c c c c c @{}}
			\multicolumn{7}{c}{Set-up 2 with two {\sl straight DoAs}}\\
			\hline
			 & 	&SAEN  &EN &Lasso  &OMP &CoSaMP \\
			\midrule
			PER    &(1.000)  &{\bf 1.000}  &0.981   &0.981 &0.998 &0.678 \\
            RMSE    &  &{\bf 0.126}   &0.145 &0.145  &0.128 &1.436 \\

            \midrule\midrule
            
			\multicolumn{7}{c}{Set-up 3 with two {\sl oblique DoAs}}\\
			\hline
			        & &SAEN  &EN &Lasso  &OMP &CoSaMP \\
			\midrule
			PER    &(1.000)  &{\bf 0.978}  &0.399   &0.399 &0.613 &0.113 \\
            RMSE    &  &{\bf 0.154}   &0.916 &0.916  &0.624 &2.296 \\

            \midrule\midrule
            
			\multicolumn{7}{c}{Set-up 4 with three {\sl oblique DoAs}}\\
			\hline
			        & &SAEN  &EN &Lasso  &OMP &CoSaMP \\
			\midrule
			PER    &(0.776)  &{\bf 0.749}  &0.392   &0.378 &{\it 0} &{\it 0} \\
            RMSE    &  &{\bf 0.505}   &0.838 &0.827  &1.087 &5.290 \\
\end{tabular*} }
\end{ruledtabular}
\end{table}

Next we discuss the results for set-up~4 which is similar to set-up~ 3, except that we have introduced a third source that also arrives from an oblique DoA  $\theta = 43^\circ$ and 
the variation of the source powers is slightly larger.  As can be noted from Table~\ref{table:sc2a3},   the PER rates of   greedy algorithms, OMP and CoSaMP,  have declined to outstandingly low 0\%.
This is very different with the PER rates they had in set-up~3 which contained only two sources. Indeed, inclusion of the 3rd source from an DoA similar with the other two sources completely ruined their accuracy.  This is in deep contrast with the SAEN method that still achieves   PER rate of 75\%,  which is more than twice the PER rate achieved by  Lasso.  SAEN is again having the lowest RMSE values.

\begin{figure}[h]
	\centering
\figline{
\fig{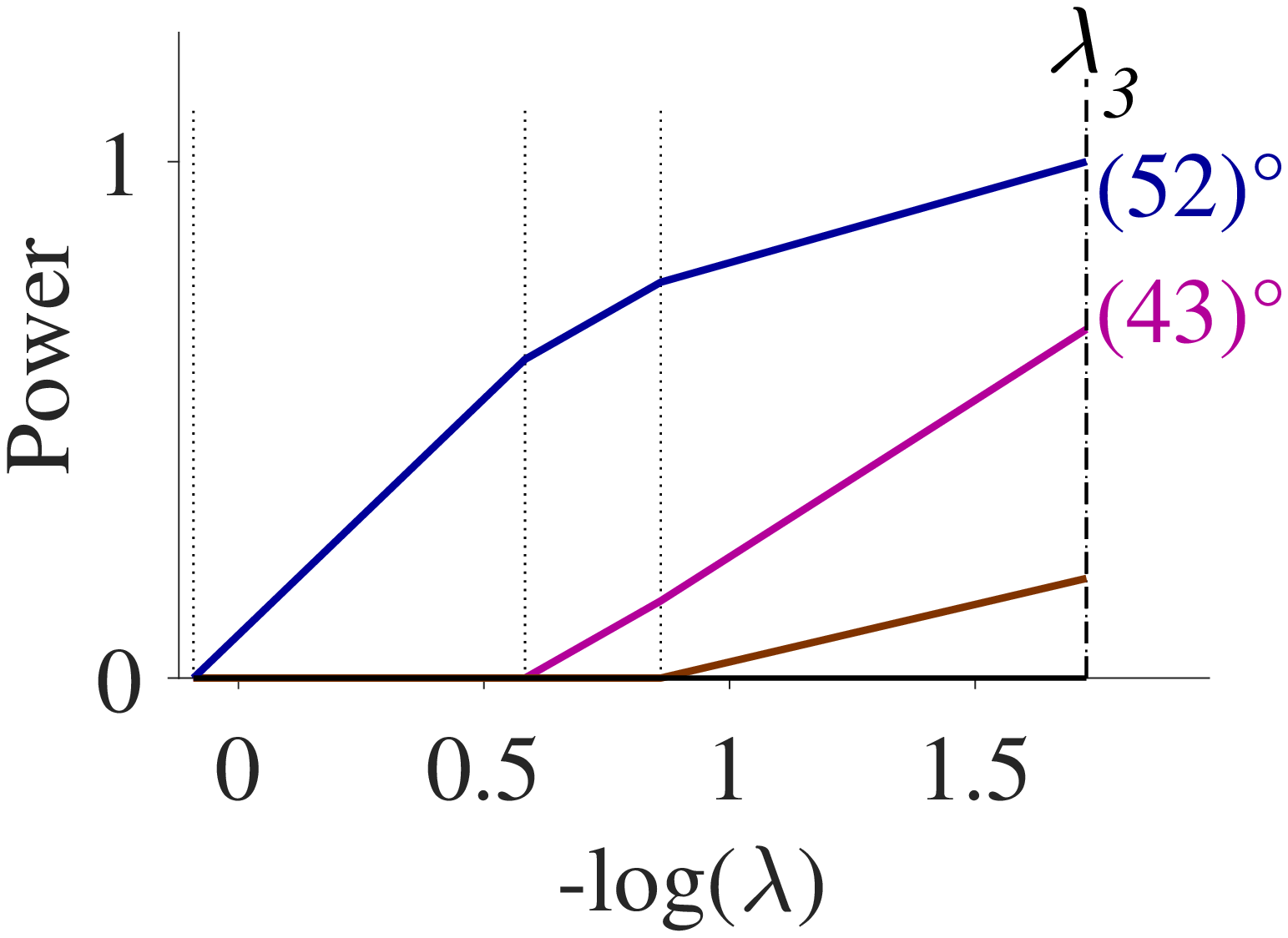}{0.51\columnwidth}{(a)}
\fig{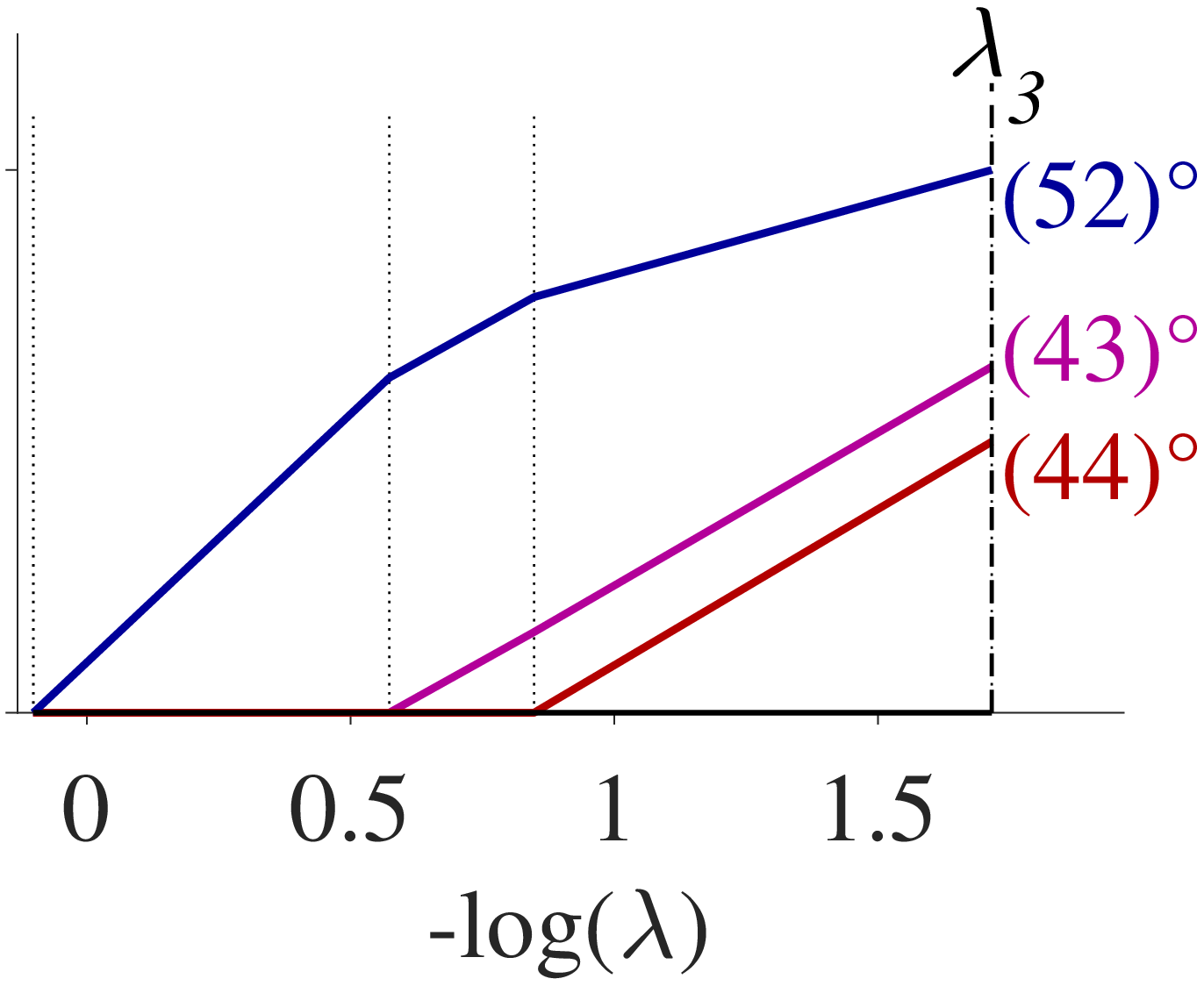}{0.45\columnwidth}{(b)}
}
\figline{
\fig{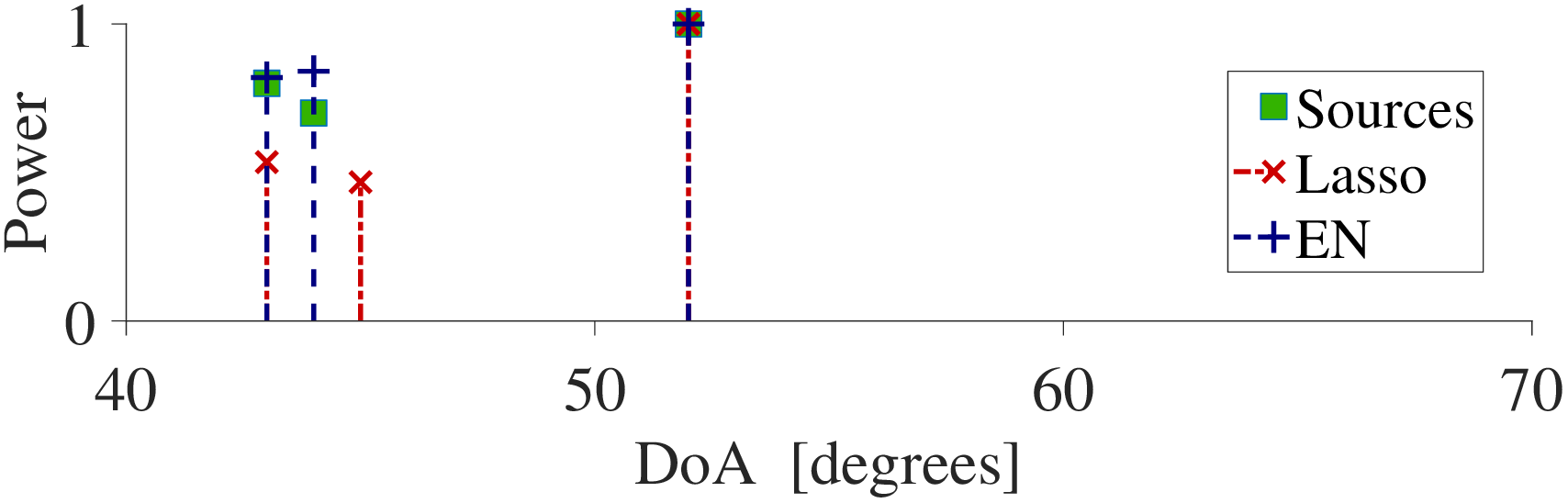}{0.98\columnwidth}{(c)}
}
\caption{(Color online) The Lasso and EN solution paths (upper panel) and respective DoA solutions at the knot $\lam_3$. Observe that  Lasso fails to recover the true support but EN successfully picks the true DoAs. The EN tuning parameter was $\al=0.9$ in this example.} \label{fig:las0en1}
\end{figure}

In summary, the recovery results for set-ups 1-4 (which express different degrees of basis coherence, proximity of target DoA's, as well as variation of source powers), clearly  illustrate  that the proposed SAEN  performs very well in  identifying the true support and  the power of the sources and is always outperforming the commonly used benchmarks sparse recovery methods, namely, the Lasso, EN,  OMP or CoSaMP with a significant margin.  
It is also noteworthy that EN often achieved better PER rates than Lasso which is mainly due to its group selection ability.  As a specific example of this particular feature,  \autoref{fig:las0en1} shows the solution paths for Lasso and EN for one particular Monte-Carlo trial, where EN correctly chooses the true DoAs but  Lasso fails to select  {\it all}  correct DoAs. In this particular instance, the EN tuning parameter was $\al=0.9$. This is reason behind the success of our c-PW-WEN algorithm which is the core computational engine of the SAEN. 

\subsection{Off-grid sources}

Set-ups 5 and 6 explore the case when the target DoAs are off the grid. Also note that set-up 5 uses finer grid spacing $\delTh = 1^\circ$ compared to set-up 6 with $\delTh = 2^\circ$. Both set-ups contain four target sources that have largely varying source powers.  In the off-grid case,  one would like the CBF method to localize the targets  to the nearest DoA in the angular grid $[\vartheta]$ that is used to construct the array steering matrix. Therefore, in the off the grid case, the PER rate refers to the case that CBF method selects the $\kdim$-sparse  support that corresponds to DoA's on the grid that are closest in distance to the true DoAs.  
Table \ref{table:sc6a8} provides the recovery results.  As can be seen, again the SAEN  is performing very well, outperforming the Lasso and EN. Note that OMP and CoSaMP completely fail in selecting the nearest grid-points.

\begin{table}[hbt]
    \setlength\extrarowheight{2pt}
	\setlength{\textfloatsep}{-0.1cm}
	\centering
	\caption{Performance results of CBF methods for set-ups 5 and 6, where target DoA-s are off the grid.  Here PER rate refers to the case that CBF method selects the $\kdim$-sparse  support that corresponds to DoA's on the grid  that are closest to the true DoAs.   The upper bound \eqref{eq:upperbound} for the PER rate of SAEN is given in parentheses. SNR level is 20 dB. } \label{table:sc6a8}
	\begin{ruledtabular}
    {\begin{tabular*}{0.97\columnwidth}{@{\extracolsep{\fill}} l | c c c c c c @{}}
			\multicolumn{7}{c}{Set-up 5 with four off-grid {\sl straight DoAs} }\\
			\hline
			 & &SAEN  &EN &Lasso  &OMP &CoSaMP \\
			\midrule
			PER    & (0.999)  &{\bf 0.649}  &0.349   &0.328 & 0 &  0 \\
            RMSE    &  &{\bf 0.899}   &0.947 &0.943  &1.137 &89.09 \\

            \midrule\midrule
            
			\multicolumn{7}{c}{Set-up 6 with four off-grid {\sl oblique DoAs} }\\
			\hline
			        &  &SAEN  &EN &Lasso  &OMP &CoSaMP \\
			\midrule
			PER    &(0.794)  &{\bf 0.683}  &0.336   &0.336 &  0 &0.005 \\
            RMSE    &  &{\bf 0.811}   &0.913 &0.911  &1.360 &28919 \\
\end{tabular*} }
\end{ruledtabular}
\end{table}

\subsection{More targets and varying SNR levels}

\begin{figure}[t]
	\centering
	\includegraphics[width=0.98\columnwidth]{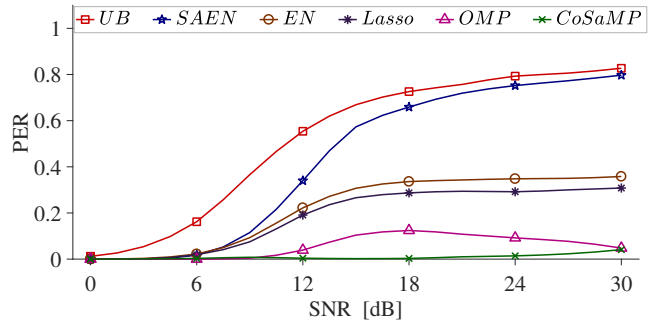}
	\caption{(Color online) PER rates of CBF methods  at different SNR levels for set-up~7.}
	\label{fig:snrVpers}
\end{figure}

Next we consider the set-up~7 (cf. Table \ref{table:scene})  which contains $K=4$  sources. The first three of the sources are  at straight DoAs and the fourth one at a DoA with modest  obliqueness ($\theta_4 = 18^o$). We now compute the PER rates of the methods as a function of SNR.   
From the PER rates shown in Fig.  \ref{fig:snrVpers}  we again notice that SAEN clearly outperforms all of the other methods.  Note that the upper bound \eqref{eq:upperbound} of the PER rate of the SAEN is also plotted.  Both greedy algorithms,  OMP and CoSaMP,  are performing very poorly even at high SNR levels.  Lasso and EN are attaining better recovery results than the greedy algorithms. Again EN is performing  better than Lasso due to additional flexibility offered by EN tuning parameter and its  ability to cope with correlated steering (basis) vectors.  
SAEN recovers the exact true support in most of the cases due to its step-wise adaptation using cleverly chosen weights.  
Furthermore, the improvement in PER rates offered by SAEN becomes larger as the SNR level increases. One can also notice that SAEN is close to the theoretical upper bound of PER rate at higher SNR regime.  

\section{Conclusions}  \label{sec:concl}

We developed c-PW-WEN algorithm that computes weighted elastic net solutions at the knots of penalty parameter over a grid of EN tuning parameter values.  c-PW-WEN also computes weighted Lasso as a special case (i.e., solution at $\al=1$) and adaptive EN (AEN) is obtained when  adaptive (data dependent) weights are used.  We then proposed a novel SAEN approach that uses c-PW-WEN method as its core computational engine and uses three-step adaptive weighting scheme where sparsity  is decreased from $3K$ to $K$ in three steps. Simulations illustrated that SAEN performs better than the  adaptive EN approaches. Furthermore, we illustrated that the $3\kdim$-sparse initial solution  computed at step 1 of SAEN provide smart weights for further steps and includes the true $\kdim$-sparse  support with high accuracy.  The proposed  SAEN algorithm is then accurately including the true support at each step.

Using the $\kdim$-sparse Lasso solution computed directly from Lasso path at the  $k^{th}$ knot fails to provide exact support recovery in many cases, especially when we have high basis coherence and lower SNR.   Greedy algorithms often fail in the face of high mutual coherence  (due to dense grid spacing or oblique target DoA's) or low SNR. This is mainly due the fact that their performance heavily depends on their ability to accurately detecting maximal correlation between the measurement vector $\y$ and the basis vectors (column vectors of $\X$). Our simulation study also showed that their performance (in terms of PER rate) deteriorates when the number of targets increases.   In the off-grid case, the greedy algorithms also failed to find the nearby grid-points. 

Finally, the SAEN algorithm performed better than all other methods in each set-up and the improvement was more pronounced in the presence of high mutual coherence.
This is due to  ability of SAEN to include the true support correctly at all three steps of the algorithm. 
Our MATLAB\raisebox{1pt}{\textregistered} package that implements the proposed algorithms is freely available at \url{www.github.com/mntabassm/SAEN-LARS}. The package also contains a MATLAB\raisebox{1pt}{\textregistered} live script demo on how to use the method in CBF problem along with an example from simulation set-up 4 presented in the paper.

 \begin{acknowledgments}
The research was partially supported by the Academy of Finland grant no. 298118 which is gratefully acknowledged.
 \end{acknowledgments}




%


\end{document}